\documentclass[journal=jacsat,manuscript=article]{achemso}

\usepackage[version=3]{mhchem}
\usepackage{physics}
\usepackage{diagbox}
\usepackage{subcaption}
\usepackage{xcolor}
\author{Zhanning Wang}
\email{zhanning.wang@unsw.edu.au}
\affiliation[The University of New South Wales]
{School of Physics, The University of New South Wales, Sydney, New South Wales 2052, Australia}
\alsoaffiliation[The University of New South Wales]{Australian Research Council Centre of Excellence in the Future Low-Energy Electronics Technologies,
The University of New South Wales, Sydney, New South Wales 2052, Australia}

\author{Elizabeth Marcellina}
\affiliation[The University of New South Wales]
{School of Physics, The University of New South Wales, Sydney, New South Wales 2052, Australia}
\alsoaffiliation[The University of New South Wales]{Australian Research Council Centre of Excellence in the Future Low-Energy Electronics Technologies,
The University of New South Wales, Sydney, New South Wales 2052, Australia}

\author{A. R. Hamilton}
\affiliation[The University of New South Wales]
{School of Physics, The University of New South Wales, Sydney, New South Wales 2052, Australia}
\alsoaffiliation[The University of New South Wales]{Australian Research Council Centre of Excellence in the Future Low-Energy Electronics Technologies,
The University of New South Wales, Sydney, New South Wales 2052, Australia}

\author{James H. Cullen}
\affiliation[The University of New South Wales]
{School of Physics, The University of New South Wales, Sydney, New South Wales 2052, Australia}

\author{Sven Rogge}
\affiliation[The University of New South Wales]
{School of Physics, The University of New South Wales, Sydney, New South Wales 2052, Australia}
\alsoaffiliation[The University of New South Wales]
{Center for Quantum Computation and Communication Technology,\\
The University of New South Wales, Sydney, New South Wales 2052, Australia}

\author{Joe Salfi}
\affiliation[University of British Columbia]{Department of Electrical and Computer Engineering,
	University of British Columbia, Vancouver, BC V6T 1Z4, Canada}

\author{Dimitrie Culcer}
\email{d.culcer@unsw.edu.au}
\affiliation[The University of New South Wales]
{School of Physics, The University of New South Wales, Sydney, New South Wales 2052, Australia}
\alsoaffiliation[The University of New South Wales]{Australian Research Council Centre of Excellence in the Future Low-Energy Electronics Technologies,
The University of New South Wales, Sydney, New South Wales 2052, Australia}
\title[\texttt{achemso} demonstration]
{Optimal operation points for ultrafast, highly coherent Ge hole spin-orbit qubits}
\begin{document}
%\keywords{Keyword1, Keyword2, Keyword3}

\begin{abstract}
Strong spin-orbit interactions make hole quantum dots central to the quest for electrical spin qubit manipulation enabling fast, low-power, scalable quantum computation. Yet it is important to establish to what extent spin-orbit coupling exposes qubits to electrical noise, facilitating decoherence. Here, taking Ge as an example, we show that group IV gate-defined hole spin qubits generically exhibit optimal operation points, defined by the top gate electric field, at which they are both fast and long-lived: the dephasing rate vanishes to first order in electric field noise along all directions in space, the electron dipole spin resonance strength is maximised, while relaxation is drastically reduced at small magnetic fields. The existence of optimal operation points is traced to group IV crystal symmetry and properties of the Rashba spin-orbit interaction unique to spin-3/2 systems. Our results overturn the conventional wisdom that fast operation implies reduced lifetimes, and suggest group IV hole spin qubits as ideal platforms for ultra-fast, highly coherent scalable quantum computing.
\end{abstract}

\section*{Introduction}
Quantum computing architectures require reliable qubit initialisation, robust single-qubit operations, long coherence times, and a clear pathway towards scaling up. Solid-state platforms are supported by the well developed solid-state device industry, with mature microfabrication and miniaturisation technologies. Among solid-state platforms, semiconductor quantum dot (QD) spin qubits have been actively pursued,\cite{Loss.PRA} with an energetic recent focus on hole spins in diamond and zincblende nano-structures.\cite{Coish.SSC.Spin.interactions.relaxation.and.decoherence.in.quantum.dots,Stefano.EPJ.Controlling.hole.spin.in.quantum.dots.and.wells,Veldhorst.PRB.Light.effective.hole.mass.in.undoped.Ge/SiGe.quantum.wells,Alex.NC.Spin.and.orbital.structure.of.the.first.six.holes.in.a.silicon.metal-oxide-semiconductor.quantum.dot,Katsaros.NC.A.germanium.hole.spin.qubit,Katsaros.NanoLetter.Heavy-Hole.States.in.Germanium.Hut.Wires,Katsaros.NanoLetter.Single-Shot.Readout.of.Hole.Spins.in.Ge,FranceschiSiGe.quantum.dots.for.fast.hole.spin.Rabi.oscillations.APL,Franceschi.NanoLetter.Ballistic,Sanquer.APL.Pauli.blockade.in.a.few-hole.PMOS.double.quantum.dot.limited.by.spin-orbit.interaction,Zwanenburg.Nanoletter.Spin.state.of.the.first.four.holes,Zwanenburg.APL.Single-hole.tunneling.through.a.two-dimensional.hole.gas.in.intrinsic.silicon,Zwanenburg.PRB.Anisotropic.Pauli.spin.blockade.in.hole.quantum.dots,Joe.PRL.Spin-Orbit.Qubit.in.Silicon,Joe.Nanotech.acceptor,Elizabeth.PRB.Spin.blockade.in.hole.quantum.dots,Terrazos.arxiv.Strained.Ge,sven.SA,Vandersypen.NM.Nuclear.spin.effects.in.semiconductor.quantum.dots,Kyrychenko.PRB.Diluted.magnetic.semiconductor.quantum.dots,Kouwenhoven.N.Spin-orbit.qubit.in.a.semiconductor.nanowire,Golovach.PRB.singlet-triplet,Kodera.APL.Detection.of.variable.tunneling.rates.in.silicon.quantum.dots,Kodera.NN.A.quantum-dot.spin.qubit.with.coherence.limited.by.charge.noise.and.fidelity.higher,PhysRevB.99.085308}

The primary motivation for this focus is the strong hole spin-orbit interaction, which enables qubit control via electron dipole spin resonance (EDSR), making quantum computing platforms faster, more power-efficient and easier to operate. \cite{Loss.PRL.EDSR.for.heavy.holes.in.quantum.dots,Golovach.PRB.EDSR,Coish.PSS.Measurement.control.and.decay.of.quantum-dot.spins,Partoens.PRB.All-electrical.control.of.quantum.gates.for.single.heavy-hole.spin.qubits,Trauzettel.PRB.All-electric.qubit.control.in.heavy.hole.quantum.dots.via.non-Abelian.geometric.phases,Ensslin.N.Electrical.control.of.spin.coherence.in.semiconductor.nanostructures,Pribiag.NN.Electrical.control.of.single.hole.spins.in.nanowire.quantum.dots,Khaetskii.PE.Spin.relaxation.in.semiconductor.mesoscopic.systems} Electric fields are much easier to apply and localise than magnetic fields used in electron spin resonance. Only a global static magnetic field is required to split the qubit levels. In addition, the $p$-symmetry of the hole wave function causes the contact hyperfine interaction to vanish, and complications involving valley degrees of freedom are absent.\cite{Ekenberg.P-type,Woods.PRB.Hole.spin.relaxation.in.quantum.dots,Hawrylak.PRB.Valence.holes.as.Luttinger.spinor.based.qubits.in.quantum.dots,Loss.PRB.Strong.spin-orbit.interaction.and.helical.hole.states.in.Ge/Si.nanowires,PhysRevB.100.085305} Initial studies indicate that hole spins may possess sufficiently long coherence times for quantum computing.\cite{Partoens.PRL.Spin-Orbit-Mediated.Manipulation.of.Heavy-Hole.Spin.Qubits.in.Gated.Semiconductor.Nanodevices,Khaetskii.PRB.Spin.relaxation.in.semiconductor.quantum.dots,Khaetskii.PRL.Electron.Spin.Decoherence.in.Quantum.Dots.due.to.Interaction.with.Nuclei,LossPRL.Hybridization.and.Spin.Decoherence.in.Heavy-Hole.Quantum.Dots,kobayashi2018engineering} Meanwhile, much progress has been made in the initialization and readout of hole spin qubits.\cite{Katsaros.NanoLetter.Single-Shot.Readout.of.Hole.Spins.in.Ge,Sanquer.APL.Pauli.blockade.in.a.few-hole.PMOS.double.quantum.dot.limited.by.spin-orbit.interaction,Elizabeth.PRB.Spin.blockade.in.hole.quantum.dots,Zwanenburg.PRB.Anisotropic.Pauli.spin.blockade.in.hole.quantum.dots,Kodera.PRB.Magnetic.field.dependence.of.Pauli.spin.blockade,hendrickx2019fast}

The existential question that will determine the future of hole QD spin qubits is: Does the strong spin-orbit interaction that allows fast qubit operation also enhance undesired couplings to stray fields such as phonons and charge noise leading to intractable relaxation and dephasing? In this paper, we demonstrate theoretically that this is emphatically not the case for hole spin qubits in group IV materials taking Ge as the most prominent example.  

In fact, the unique properties of the hole Rashba interaction overturn the conventional understanding of qubit coherence in spin-orbit coupled systems, which states that, as long as the qubit is described by an effective $2 \times 2$ Hamiltonian, holes behave in the same way as electrons. That is, given that the EDSR rate is linear in the spin-orbit strength, while the relaxation and dephasing rates increase as the square of the spin-orbit strength, the Rabi ratio can be enhanced by operating the qubit at points at which the spin-orbit interaction is weaker. \cite{Golovach.PRB.EDSR,Loss.PRL.EDSR.for.heavy.holes.in.quantum.dots} In contrast group IV hole qubits achieve the best coherence when the electrical driving rate is at its maximum. In all other systems one has to choose between creating long-lived qubits by isolating them from the environment and accepting slower gate times, or designing fast qubits strongly coupled to the environment, but which decohere rapidly.

The key realisation is that holes in group IV materials are qualitatively different from group III-V materials. They have tremendous potential for qubit coherence, with Ge and Si possessing isotopes with no hyperfine interaction as well as a near-inversion symmetry that eliminates piezo-electric phonons. This near-inversion symmetry will eliminates the Dresselhaus interaction, the interface inversion asymmetry terms are expected to be negligible in the systems.\cite{PhysRevB.89.075430,Elizabeth.PRB.variational.analysis} Which enable spin resonance in group III-V materials \cite{Loss.PRL.EDSR.for.heavy.holes.in.quantum.dots}. On the other hand we show that strong cubic-symmetry terms enable a fully-tunable Rashba coupling. Unlike the Dresselhaus interaction, and unlike electron systems, the hole Rashba interaction evolves nonlinearly as a function of the gate electric field, a fact traced to the hole spin-3/2. The qualitative difference between the Rashba and Dresselhaus interactions for holes is vital for qubit coherence. Thanks to this nonlinearity, dephasing due to electric field fluctuations in all spatial directions can be essentially eliminated at specific optimal operation points defined by the gate electric field. \cite{Joe.Nanotech.acceptor, Joe.PRL.Spin-Orbit.Qubit.in.Silicon,Adam.APL.dephasing,Culcer.APL.Si.charge.noise.,CoishPRB.Spin.decoherence.of.a.heavy.hole.coupled.to.nuclear.spins.in.a.quantum.dot,Zulicke.PRB.spin-3/2,Loss.PRB.Strong.spin-orbit.interaction.and.helical.hole.states.in.Ge/Si.nanowires,Ensslin.NP.Measurement.of.Rashba.and.Dresselhaus.spin-orbit.magnetic.fields,Zulicke.PRB.Spin.accumulation.in.quantum.wires.with.strong.Rashba.spin-orbit.coupling} At these points electrical qubit rotations are at their most efficient, with the spin resonance Rabi gate time at a minimum. The relaxation rate due to phonons can be made as small as desired at small magnetic fields of the order of $0.1\,\text{T}$, which allows $10^6-10^7$ operations in one relaxation time for an in-plane alternating field $E_{\text{AC}} \sim 10^3\,\text{V}\text{m}^{-1}$. 

Our focus in this paper is on Ge, which has witnessed enormous recent progress.  \cite{FranceschiSiGe.quantum.dots.for.fast.hole.spin.Rabi.oscillations.APL,Katsaros.NC.A.germanium.hole.spin.qubit,Katsaros.NanoLetter.Heavy-Hole.States.in.Germanium.Hut.Wires,Franceschi.NanoLetter.Ballistic,guoguoping.APL.Measuring.hole.spin.states.of.single.quantum.dot.in.germanium.hut.wire} Holes in planar Ge quantum wells have a very large out of plane Land\'e g-factor, $g \approx 20$, enabling operation at very small magnetic fields, which would not impede coupling to a superconducting resonator. The low resistivity of Ge when contacting with metals makes couplings between other devices such as superconductors easier.\cite{guoguoping.NanoLetter.Coupling.a.Germanium.Hut.Wire.Hole.Quantum.Dot.to.a.Superconducting.Microwave.Resonator,FranceschiSiGe.quantum.dots.for.fast.hole.spin.Rabi.oscillations.APL,Loss.PRB.cQED} In the past decade some spectacular results have been reported, for example, EDSR detection techniques,\cite{Terrazos.arxiv.Strained.Ge, guoguoping.APL.Measuring.hole.spin.states.of.single.quantum.dot.in.germanium.hut.wire} structures of quantum confinement systems,\cite{Giordano.Advanced.Functional.Materials, Hardy.Nanotech.Single.and.double.hole.quantum.dots.in.strained.Ge.SiGe.quantum.wells, Kodera.APL.Characterization.and.suppression.of.low-frequency.noise.in.Si.SiGe.quantum.point.contacts.and.quantum.dots, Katsaros.NanoLetter.Heavy-Hole.States.in.Germanium.Hut.Wires} the anisotropy of $g$-tensors,\cite{Franceschi.NanoLetter.Ballistic,Giordano.Advanced.Functional.Materials} spin-orbit couplings and transport phenomena in two-dimensional hole systems.\cite{Veldhorst.PRB.Light.effective.hole.mass.in.undoped.Ge/SiGe.quantum.wells, Katsaros.NC.A.germanium.hole.spin.qubit, Franceschi.NanoLetter.Ballistic, Katsaros.PRL.Observation.of.Spin-Selective.Tunneling.in.SiGe.Nanocrystals,Rol_2019}
We focus a on single dot throughout this work. A prototype device, including a neighboring dot, is shown in Figure 1. The Hamiltonian describing a single hole quantum dot has the general form $H = H_{\text{LK}}+ H_{\text{BP}}+ H_{\text{Z}} + H_{\text{ph}}+H_{\text{conf}} $, where $H_{LK}$ represents the Luttinger-Kohn Hamiltonian, $H_\text{Z}$ is the Zeeman interaction between the hole and an external magnetic field, and $H_{ph}$ the hole-phonon interaction. $H_{\text{conf}}$ is the confinement potential including the vertical and lateral confinement. The vertical confinement is achieved by applying a gate electric field $F_\text{z}$ in the growth direction, leading to a term $eF_\text{z}z$ in the Hamiltonian; the lateral confinement is modelled as an in-plane parabolic potential well. The Bir-Pikus Hamiltonian $H_{\text{BP}}$ represents strain, \cite{Terrazos.arxiv.Strained.Ge} which appears naturally as part of the quantum well growth process. A typical configuration of holes in Ge is achieved by growing a thin strained Ge layer (usually about $10\,\text{nm}$ to $20\,\text{nm}$) between SiGe layers such that, if the barrier between the two layers is high enough, a quantum well can be formed. We consider $\text{Si}_{x}\text{Ge}_{1-x}$, where $x=0.15$, \cite{Katsaros.NC.A.germanium.hole.spin.qubit,Katsaros.NanoLetter.Heavy-Hole.States.in.Germanium.Hut.Wires,Veldhorst.PRB.Light.effective.hole.mass.in.undoped.Ge/SiGe.quantum.wells,Giordano.Advanced.Functional.Materials}
with other values of $x$ discussed in the Supplementary Table 1. 

We start from the bulk band structure of holes as derived by Luttinger and Kohn.\cite{Luttinger} The spinor basis is formed by the eigenstates of $J_\text{z}$, $\left\{\left|+\frac{3}{2}\right\rangle, \left|-\frac{3}{2}\right\rangle, \left|+\frac{1}{2}\right\rangle, \left|-\frac{1}{2}\right\rangle\right\}$. For a 2D hole gas grown along $\hat{z} \parallel (001)$, we write the Luttinger-Kohn Hamiltonian as:
\begin{equation}\label{eq:Luttinger_Kohn_Hamiltonian_matrix}
    H_{\mathrm{LK}}\left(k^{2}, k_\text{z}\right)=\left[\begin{array}{cccc}{P+Q} & {0} & {L} & {M} \\ {0} & {P+Q} & {M^{*}} & {-L^{*}} \\ {L^{*}} & {M} & {P-Q} & {0} \\ {M^{*}} & {-L} & {0} & {P-Q}\end{array}\right],
\end{equation}
where $P = \frac{\hbar^2\gamma_{1}}{2m_0} \left(k^{2}+k_{\text{z}}^{2}\right)$, $Q = -\frac{\hbar^2\gamma_{2}}{2m_0} \left(2 k_\text{z}^{2}-k^{2}\right)$, $L = -\sqrt{3} \frac{\hbar^2\gamma_{3}}{2m_0} k_{-} k_\text{z}$, $M = -\frac{\sqrt{3}\hbar^2}{2m_0}\left(\overline{\gamma} k_{-}^{2}+\delta k_{+}^{2}\right)$, and $m_0$ is the free electron mass, $\gamma_1$, $\gamma_2$, $\gamma_3$ are Luttinger parameters which are determined by the band structure. The in-plane wave vector will be $k^{2}=k_{x}^{2}+k_{y}^{2}$, $k_{\pm}=k_{x} \pm i k_{y}$. In this manuscript we use the symmetric gauge $\textbf{\text{A}}=(B / 2)(-y, x, 0)$, so that $\textbf{\text{k}} \rightarrow -i \boldsymbol{\nabla} + e\textbf{\text{A}}/\hbar$. We have also used $\bar{\gamma} = (\gamma_2 + \gamma_3)/2$ and $\delta = (\gamma_3 - \gamma_2)/2$ to simplify the algebra. In Ge $\delta/\bar{\gamma} < 0.15 $, hence $\delta$ can be treated perturbatively, while bulk Dresselhaus terms are absent. Although interface inversion asymmetry terms with the same functional form may exist,\cite{PhysRevB.89.075430} at the strong gate fields considered here they will be overwhelmed by the Rashba interaction and are not discussed in detail.
The diagonal terms of $H_{\text{BP}}$ in the HH manifold are $P_\varepsilon+Q_\varepsilon=-a_\text{v}\left(\varepsilon_{xx}+\varepsilon_{yy}+\varepsilon_{zz}\right)$, while in the LH manifold they are  $P_\varepsilon-Q_\varepsilon=-(b_\text{v}/2)\left(\varepsilon_{\text{xx}}+\varepsilon_{\text{yy}}-2\varepsilon_{\text{zz}}\right)$, where $a_\text{v}=-2\,\text{eV}$ and $b_\text{v}=-2.3\,\text{eV}$ are deformation potential constants.\cite{Terrazos.arxiv.Strained.Ge} In our chosen configuration $\varepsilon_{\text{xx}}=\varepsilon_{\text{yy}}=-0.006$, the minus sign indicates that the germanium is compressed in $xy$-plane. In the $\hat{\text{z}}$-direction, the Ge layer will be stretched, and $\varepsilon_{\text{zz}}=(-2C_{12}/C_{11})\varepsilon_{\text{xx}}=0.0042$, with $C_{12}=44\,\text{GPa}$, $C_{11}=126\,\text{GPa}$ for Ge. The diagonal terms of the strain-relaxed barrier configuration will change the HH-LH energy splitting by a constant, which is approximately $50\,\text{meV}$. 

The growth direction provides the spin quantization axis, with the heavy holes states (HHs) representing the $\pm3/2$ angular momentum projection onto this axis, while the light holes states (LHs) represent $\pm1/2$. In 2D hole systems, the HHs are the ground state. \cite{Loss.PRL.EDSR.for.heavy.holes.in.quantum.dots,Katsaros.NanoLetter.Heavy-Hole.States.in.Germanium.Hut.Wires,CoishPRB.Spin.decoherence.of.a.heavy.hole.coupled.to.nuclear.spins.in.a.quantum.dot,Elizabeth.PRL.Zeeman.Spin.Splitting.in.Two-Dimensional.Hole.Systems,Hong.PRL} To define a quantum dot a series of gates are added on top of the 2D hole gas confinement, as in Figure 1, and we ultimately seek an effective Hamiltonian describing the two lowest-lying HH states in a quantum dot. Since we expect the HH-LH splitting to be much larger than the quantum dot confinement energy, we proceed with the standard assumptions of ${\textbf{\text{k}}}\cdot{\textbf{\text{p}}}$ theory, retaining at first only terms containing $k_\text{z}$, with $k_\text{x}$ and $k_\text{y}$ initially set to zero. This determines the approximate eigenstates $\psi_{\text{H},\text{L}}(z)$ corresponding to the growth-direction. These are described by two variational Bastard wave functions $\psi_\text{H}$ and $\psi_\text{L}$, \cite{Bastard,Elizabeth.PRB.variational.analysis}
\begin{equation}\label{eq:Bastard_variation_function}
   \psi_{\text{H},\text{L}}(z)=\sqrt{\frac{4 \beta_{\text{H},\text{L}} \left(\pi^{2}+\beta_{\text{H},\text{L}} ^{2}\right)}{\left(1-e^{-2 \beta_{\text{H},\text{L}} }\right) d \pi^{2}}} \cos \left(\frac{\pi z}{d}\right) \exp \left[-\beta_{\text{H},\text{L}} \left(\frac{z}{d}+\frac{1}{2}\right)\right],
\end{equation}
where the dimensionless variational parameters $\beta_{\text{H},\text{L}}$ are sensitive to the gate electric field due to the term $eF_\text{z}z$, and $d$ is the width of the quantum well in the growth direction, which is an input parameter. The orthogonality of the HH and LH states is ensured by the spinors. This wave function is suitable for inversion layers as well as accumulation layers, although our focus will be primarily on the latter. For inversion layers, the Bastard wave function will also be appropriate, because in experiments the electric field can be made large enough such that the hole gas sticks around the top of the quantum well.

In the $xy$-plane, we model the confinement potential using a harmonic oscillator
\begin{equation}
	\left[\frac{\hbar^{2}}{2 m_{p}}\left(-i \nabla_{\|}+e \textbf{\text{A}}\right)^{2}+\frac{1}{2} m_{\text{p}} \omega_{0}^{2}\left(x^{2}+y^{2}\right)\right] \phi=\varepsilon \phi
\end{equation}
where $m_\text{p}=m_0/(\gamma_1+\gamma_2)$ is the in-plane effective mass of the heavy holes, the subscript $\parallel$ refers to the $xy$-plane, $\omega_0$ is the oscillator frequency, $a_0$ the QD radius which satisfy $a_0^2=\hbar/(m_\text{p}\omega_\text{l})$, i.e. a magnetic field will narrow the QD radius. The solutions are the well-known Fock-Darwin wave functions $\ket{\phi_{n_1,n_2}}$\cite{Terrazos.arxiv.Strained.Ge}
with eigen-energies $\varepsilon_{n_1,n_2}=\hbar(n_1+n_2+1)\omega_l + \frac{1}{2}\hbar(n_2-n_1)\omega_\text{c}$, where $\omega_\text{l}=\sqrt{\omega_0^2+\omega_\text{c}^2/4}$,  $\omega_\text{c}=eB/m_\text{p}$ is the cyclotron frequency. The Bastard wave functions account for the perpendicular confinement, in the $\hat{\textbf{\text{z}}}$-direction. The Fock-Darwin wave functions $\ket{\phi_{n_1,n_2}}$ account for the quantum dot confinement in the $xy$-plane. The Bastard and Fock-Darwin wave functions would be formally the same for electrons, while in a hole gas is that separate Bastard wave functions are required for the heavy and light holes.

Finally, the hole-phonon interaction is: \cite{Golovach.PRL.Phonon-Induced.Decay.of.the.Electron.Spin.in.Quantum.Dots, Trif.PRL.Relaxation.of.Hole.Spins.in.Quantum.Dots.via.Two-Phonon.Processes, Loss.PRL.Spin.Relaxation.and.Decoherence.of.Holes.in.Quantum.Dots}
\begin{equation}\label{eq:phonon_hole_interactions}
    H_{\text{i},\text{j},\text{s}}=\sum_{\alpha, \beta=\text{x}, \text{y}, \text{z}} \frac{1}{2} \sqrt{\frac{\hbar}{2 {N} {V}_\text{c} \rho \omega_{\text{s}}}} D_{\alpha, \beta}^{\text{i}, \text{j}}\left[\frac{q_{\alpha} \hat{{e}}_{\text{s}, \beta}+q_{\beta} \hat{{e}}_{\text{s}, \alpha}}{q}\right] q \left(e^{-i \textbf{\text{q}} \cdot \textbf{\text{r}}} \hat{a}_{q}^{\dagger}+e^{i \textbf{\text{q}} \cdot \textbf{\text{r}}} \hat{a}_{q}\right),
\end{equation}
where $q$ is the phonon wave vector, ${V}_\text{c}$ is the unit cell volume, $NV_\text{c}$ is the crystal volume, $\hat{e}_\text{s}$ is the polarization direction vector. The density of the material is denoted by $\rho$, $D_{\alpha,\beta}$ represents the deformation potential matrix, and $\hat{a}^\dagger$ and $\hat{a}$ are the phonon creation and the annihilation operators. The details are presented in the Supplementary Information Eq.(1).\cite{Loss.PRB.g-factor.phonon-mediated, Woods.PRB.Spin.relaxation.in.quantum.dots, Woods.PRB.Hole.spin.relaxation.in.quantum.dots, Climente.NJP.GaAs, Loss.PRB.Phonon-mediated.singlet-triplet, Golovach.PRL.Nature.of.Tunable.Hole.g.Factors.in.Quantum.Dots} 

 Our approach is semi-analytical. To incorporate the contributions from the LHs, and the Rashba spin-orbit couplings, we start from the $4 \times 4$ Luttinger Hamiltonian and project it onto the product states of the out-of-plane sub-bands HH1, LH1, each with two spin projections, and the first four orbital levels of the in-plane confinement, such that our Hamiltonian matrix is $40 \times 40$. This refers to states $\left\{ \ket{\frac{3}{2},S}\otimes\ket{\phi_{n,m}} \right\}$, where $S = 3/2, -3/2, 1/2, -1/2$ represent the HHs or LHs, $n=0,1,2,3$ and $m= 0,\pm1,2,\pm3$ denote the in-plane Fock-Darwin states. We have checked that addition of the HH2, LH2 sub-bands does not modify the results, this is attributed to the significant energy gaps separating them from HH1, LH1. Given the high computational cost of adding these sub-bands, we have not taken them into account in the results presented here.  To obtain the matrix elements required for the dephasing and relaxation times, as well as for the EDSR Rabi frequency, we perform a 3rd order Schrieffer-Wolff transformation on the $40 \times 40$ Hamiltonian. This transformation takes into account all the spin-orbit terms that are not separable in the spatial coordinates, which are precisely the terms leading to the Rashba interaction. To ensure the accuracy of the Schrieffer-Wolff method we compare the results for the Larmor frequency with a full numerical diagonalisation of the $40 \times 40$ matrix.

%=============================================
\section*{Results and Discussion}
%=============================================
\subsection*{Qubit Zeeman splitting}
The qubit Larmor frequency has been plotted in Figure 2 as a function of the gate electric field. The Schrieffer-Wolff method agrees well with the numerical diagonalization: the location of the optimal operation point differs by only $2\%$ in the two approaches. We note the non-monotonic behaviour as a function of gate field, leading to an optimal operation point in the range 30-50 MV/m. Electric fields of such magnitude are used routinely in quantum computing experiments \cite{Katsaros.NanoLetter.Heavy-Hole.States.in.Germanium.Hut.Wires, hendrickx2019fast, Lauchte1500022, doi:10.1021/acs.nanolett.5b02561}. The non-monotonic behaviour is directly related to the behaviour of Rashba spin-orbit coupling discussed below. We note that the spatial dimensionality of the qubit is determined by the relevant energy scales, namely the $\hat{\textbf{\text{z}}}$-sub-band energy spacing compared to the energy splitting of the lateral wave functions. The heavy hole - light hole splitting, given by the perpendicular confinement, is many times larger than the in-plane qubit confinement energy, determined by the in-plane confinement, so the system is in the quasi-2D limit. Nevertheless, our findings, such as trends with the top gate field, can be interpreted qualitatively by analogy with the Rashba interaction in the asymptotic 2D limit $d \rightarrow 0$, \cite{BRAVYI2011SW} which we also obtain from Eq.\ref{eq:Luttinger_Kohn_Hamiltonian_matrix}. For a system with cubic symmetry this contains two terms with different rotational properties:
\begin{equation}\label{eq:Rashba_spin_orbit_couplings}
H_{\mathrm{SO}}=i \alpha_{2}\left(k_{+}^{3} \hat{\sigma}_{-}-k_{-}^{3} \hat{\sigma}_{+}\right)+i \alpha_{3}\left(k_{+} k_{-} k_{+} \hat{\sigma}_{+}-k_{-} k_{+} k_{-} \hat{\sigma}_{-}\right),
\end{equation} 
where $\hat{\sigma}_\pm \equiv (\hat{\sigma}_\text{x} \pm i\hat{\sigma}_\text{y})/2$. The coefficients are evaluated as: 
\begin{align}\label{eq:Rashba_spin_orbit_coupling_cubic}
\alpha_{2}&=\frac{3}{2} \frac{\mu^{2} \overline{\gamma} \gamma_{3}}{E_{\text{H}}-E_{\text{L}}}\langle \psi_\text{H} | \psi_\text{L}\rangle\left[\left\langle \psi_\text{H}\left|\hat{k}_\text{z}\right| \psi_\text{L}\right\rangle -\left\langle \psi_\text{L}\left|\hat{k}_\text{z}\right| \psi_\text{H}\right\rangle \right] \\\label{eq:Rashba_spin_orbit_coupling_linear}
\alpha_{3}&=\frac{3}{2} \frac{\mu^{2} \delta \gamma_{3}}{E_{\text{H}}-E_{\text{L}}}\langle \psi_\text{H} | \psi_\text{L}\rangle\left[\left\langle \psi_\text{H}\left|\hat{k}_\text{z}\right| \psi_\text{L}\right\rangle-\left\langle \psi_\text{L}\left|\hat{k}_\text{z}\right| \psi_\text{H}\right\rangle \right],
\end{align}
where $E_\text{H}$ and $E_\text{L}$ are the energies of the lowest-lying HH and LH states, respectively, and are strong functions of the gate electric field. These formulas explain three main features. 

Firstly, the optimal operation point reflects the interplay of the quadrupole degree of freedom with the gate electric field unique to spin-3/2 systems. The behaviour of the qubit Zeeman splitting and Rashba coefficients is understood by recalling that the Rashba effect for the HH sub-bands is primarily driven by the off-diagonal matrix element $L$ in Eq.\ref{eq:Luttinger_Kohn_Hamiltonian_matrix} connecting the HH and LH sub-bands. This term, which is $\propto k_\text{z} k_+$, increases with the top gate field. At small gate fields, the Rashba spin-orbit constants increase monotonically due to the increase in the $k_\text{z}$ overlap integral. This continues until a critical top gate field is reached at which the HH-LH splittings, determined by the matrix element $Q$, begin to increase faster than the off-diagonal matrix element $L$. The heavy hole-light hole splitting induced by the confinement potential and the gate electric field is traced to the different effective masses for heavy and light holes. This physics has been shown previously by Winkler and collaborators \cite{Winkler.PRB.Rashba2000, Winkler.RPL.Highly.Anisotropic.2000, Habib.APL.Negative.differential.Rashba}. Beyond this critical field the Rashba terms decrease, resulting in a relatively broad optimal operation region at which the qubit is insensitive to background electric field fluctuations in the $\hat{z}$-direction and the dephasing rate vanishes to first order in the $\hat{z}$-electric field. As we show below, electric field fluctuations in the $\hat{z}$-direction are by far the most damaging to the qubit, and are the key source of decoherence to be avoided. The breadth and smoothness of the extreme make the tuning of the electric field to reach the optimal operation point easier.

Secondly, the sweet spot shifts slightly with the dot radius,\cite{Elizabeth.PRL.Zeeman.Spin.Splitting.in.Two-Dimensional.Hole.Systems}
and this is fully captured by our Schrieffer-Wolff results. The reason for this is that, in a 2D system, while the angular form of the Rashba interaction is dictated by rotational symmetry, the Rashba coupling constants $\alpha_2$ and $\alpha_3$ are functions of the magnitude of the wave vector. This implies they are functions of the density, and this is vital at large densities. Hence a quantum dot can be envisaged as having Rashba parameters that are functions of the in-plane radius $a_0$, and their dependence is less pronounced at larger $a_0$, since this corresponds to smaller densities.

Thirdly, each of the two spin-orbit coupling terms in Eq.\ref{eq:Rashba_spin_orbit_couplings} can be envisaged as the interaction of the hole spin with an effective spin-orbit field that depends on the momentum. In the absence of a magnetic field, the $\alpha_2$-Rashba spin-orbit field winds around the Fermi surface three times, whereas the $\alpha_3$-Rashba spin-orbit field winds only once.  {In the strict 2D limit it is the $\alpha_3$-term that enables EDSR. Although the quantum dot is not in the exact 2D limit, it still holds that EDSR is enabled by the cubic symmetry terms $\propto \delta$. Setting $\delta = 0$ in our calculations causes the EDSR frequency to vanish.}
\begin{equation}
f_{\mathrm{EDSR}}=24 g_{0} \mu_{\text{B}} B e E_{\mathrm{AC}} a_{0}^{2} \delta m_{\mathrm{p}}^{2} \times u
\end{equation}
where $u$ are long expression that can be found in the Supplementary Information Eq.(17).
\subsection*{Dephasing time}
The main dephasing mechanisms are fluctuating electrical fields such as charge noise. We focus on random telegraph noise (RTN) due to charge defects, noting that a similar discussion can be presented for $1/f$ noise, which is typically caused by an incoherent superposition of RTN sources. For this reason, we expect the trends for the two types of noise to be similar, while reliable numbers for $1/f$ noise must await experimental determination of the noise spectral density $S(\omega)$ for hole qubits. To begin with, we estimate the dephasing time $T_2^*$, which is expected to be primarily determined by fluctuations in the Larmor frequency of the qubit induced by charge noise. The electric potential induced at the qubit by a defect located at ${\textbf{\text{r}}}_\text{D}$, which may give rise to RTN, can be modelled as a quasi-2D screened Coulomb potential:
\begin{equation}
U_{\text{scr}}=\frac{e^2}{2\epsilon_0 \epsilon_\text{r}} \int_{0}^{2k_\text{F}} \frac{e^{-i {\textbf{\text{q}}} \cdot \left({\textbf{\text{r}}}-{\textbf{\text{r}}}_\text{D}\right)}}{q+q_{\text{TF}}} \frac{\dd[3]{q}}{\left(2\pi\right)^3},
\end{equation}
where $\epsilon_0$ is the vacuum permeability, $\epsilon_\text{r}$ is the relative permeability for Ge, $q_{\text{TF}}$ is the Thomas-Fermi wave vector, and $k_F$ is the Fermi wave vector.\cite{Adam.APL.dephasing} In a dilution refrigerator, the high energy modes of the Coulomb potential are negligible, therefore the $q>2k_F$ part is ignored. Another source of dephasing is dipole defects due to the asymmetry in bond polarities.
\begin{equation}
U_{\text{dip}}\left(\textbf{\text{R}}_\text{D}\right)=\frac{\textbf{\text{p}} \cdot \textbf{\text{R}}_{\text{D}}}{4 \pi \epsilon_{0} \epsilon_{\text{r}} R_{\text{D}}^{3}},
\end{equation}
where $\textbf{\text{R}}_\text{D}$ is the distance between the dot and the unscreened charge dipole. $\boldsymbol{p}$ is the dipole moment of the charge $\textbf{\text{p}} = e \textbf{\text{l}}$, the size of the dipole is about $1$\AA.

As a worst-case estimate of the dephasing time, we use the motional narrowing result,\cite{Culcer.APL.Si.charge.noise.,Adam.APL.dephasing} the dephasing time $T^{*-1}_2 = (\delta \omega)^2 \tau/2$, where $\delta\omega$ is the change in qubit Larmor frequency due to the fluctuator, and we consider $\tau=10^3t_{\text{Rabi}}$, where $t_{\text{Rabi}}$ is the single-qubit operation time (the inverse of the EDSR frequency), which can be found from Figure 3a. Because of the weak coupling between the spin degree of freedom and external reservoirs, slower fluctuators can be eliminated via pulse sequences and the spin echo techniques.\cite{press2010ultrafast} We consider two sample defects separately. One is a single-charge defect located $100\,\text{nm}$ away from the quantum dot in the plane of the dot as a worst-case scenario for a charge trap. We use $r_\text{D}=100\,\text{nm}$ since regions inside this range will be depleted by the top gate, and charge traps will not be active. We also consider a dipole defect immediately under the gate and above the dot, with ${R}_\text{D} = 20\,\text{nm}$ in the $\hat{z}$ direction. This is because within the depleted region the most relevant defects are charge dipoles, whose orientation fluctuates. To estimate the pure dephasing time at the optimal operation point due to such a defect, we first note that the in-plane electric field will not contribute to dephasing. An in-plane electric field enters the QD Hamiltonian as $ \textbf{\text{E}}_\parallel \cdot \textbf{\text{r}}_\parallel$. This in-plane electric field term does not couple states with different spin orientations. When we consider the qubit Zeeman splittings, the corrections to the effective quantum dot levels due to the in-plane electric field will read the same for $H_{1,1}$ and $H_{2,2}$ up to the second-order, therefore, fluctuations in qubit Zeeman splitting $H_{1,1}-H_{2,2}$ will not depend on the in-plane electric field. However, higher-order terms in the expansion of the electrostatic potential of the defects will lead to dephasing, and these are responsible for dephasing at the optimal operation point itself. To determine their effect, we write the ground state energy as $E_{\text{LK}} + E_0+E_\text{z}+v_0$ where $E_0$ is the lateral confinement energy, $E_\text{z}$ is the Zeeman energy, and $v_0$ is the energy correction due to the defect.

We estimate the approximate qubit window of operation around the optimal operation point. Away from the optimal operation point, due to the fluctuating electric potential of the defect, the energy levels of the quantum dot will gain a correction, i.e., $\mel**{\phi_{n_1,m_1}}{U_{\text{sc}}}{\phi_{n_2,m_2}}$. With these assumptions, the dephasing time is plotted as a function of the gate electric field in Figure 3. At the optimal operation point, the dephasing time due to the out-of-plane fluctuations is calculated to the second-order, since the first-order fluctuation vanishes, the in-plane fluctuations will dominate the dephasing. Away from the optimal operation point, the motional narrowing result is much smaller than the quasi-static limit result. This is because the first-order variation of the qubit Zeeman splitting will weaken the correlation time, while the quasi-static limit does not consider any correlations. However, as the gate electric field approaches to the optimal operation point, the variation of qubit Zeeman splitting decreases; at the optimal operation point, compared with the quasi-static limit result, longer correlation time will lead to a larger dephasing time. We also determine the pure dephasing time in the quasi-static limit, where the switching time is the longest time scale in the system. This is essentially given by $T_2=2\pi / (\delta \omega)$, and is plotted in Figure 3b. 

\subsection*{Relaxation time and EDSR}
We briefly discuss electrically driven spin resonance. An in-plane oscillating electric field represented in the Hamiltonian by $eE_{\text{AC}}(t)x$ drives spin-conserving transitions between the QD states. For a multiple occupied hole dot the excited state structure may be more complex but the argument above remains valid because the $\alpha_2$ and $\alpha_3$ Rashba terms couple the ground state to different excited states. The spin resonance Rabi time is the time taken to accomplish an operation. The Rabi frequency can be tuned by changing the gate electric field and with it the Rashba spin-orbit coupling constant. However, note that because the two Rashba terms directly determine the correction to the $g$-factor, the Rashba interaction and the $g$-factor cannot be tuned independently at present.

Given that the spin resonance frequency is a maximum at the optimal operation point, it follows that the qubit can be tuned to have maximum coherence and maximum electrical driving simultaneously. The nonlinearity in the hole Rashba interaction as a function of the gate field that enables this feature has no counterpart in electron systems. In GaAs hole systems \cite{Loss.PRL.EDSR.for.heavy.holes.in.quantum.dots} this nonlinearity does not lead to optimal operation points. This is because, firstly, the spin resonance Rabi frequency in GaAs hole qubits is driven by the Dresselhaus interaction, which is not tunable via the gate electric field, while $\alpha_3$ is negligible in GaAs. Secondly, GaAs qubits are exposed to decoherence through the hyperfine interaction, piezoelectric phonons, and the Dresselhaus interaction, none of which can be mitigated.

Since the Rabi frequency is maximised at the optimal operation point, the relaxation time $T_1$ is minimised there. For the qubit to be operated efficiently it is vital to determine the ratio of the EDSR and relaxation rates. Hyperfine interactions and phonon-hole interactions are two major factors affecting the relaxation time, hence the quality of the qubit. However, the $p$-type symmetry of the valence band excludes the contact hyperfine interaction. There is no bulk inversion asymmetry in group IV elements; this leads to no Dresselhaus spin-orbit coupling. However, there is still the Rashba spin-orbit coupling due to the structure inversion asymmetry, which couples the heavy-hole states to the light-hole states. Neither the spin nor the orbit angular momentum will be a good quantum number, as the admixture of the spin-down and the spin-up states will modify the wave functions. We emphasize that, whereas EDSR comes only from the $\alpha_{3}$-Rashba term, the qubit relaxation is caused by both the $\alpha_{2}$- and the $\alpha_{3}$-Rashba terms. 

The relaxation time evaluated using Fermi's golden rule is shown in Figure 4. For completeness, we also consider two-phonon relaxation processes, which include virtual emission and absorption of a phonon between two heavy hole states, since in the first-order relaxation calculation there is no direct matrix element between the two heavy-hole states. However, the two-phonon process calculation returns a negligible relaxation rate, which will not contribute significantly to the relaxation time. The relaxation rate will depend on the external magnetic field as $ (1/T_1) \propto B^{7}$ for the $\alpha_3$-Rashba term and $ (1/T_1) \propto B^{9}$ for the $\alpha_2$-Rashba term. This is shown in Figure 4a. 

We also plot the ratio between the relaxation time and the EDSR time, demonstrating that the system allows for a large number of operations. The allowable number of single qubit operations is calculated by evaluating the ratio of the relaxation time and the EDSR time, i.e., the Rabi ratio. The in-plane electric field we used is $E_{\text{AC}}=10^3\,\text{V/m}$. In Figure 4, we plot the relaxation time, EDSR Rabi time, comparison of the magnitude of the relaxation time and EDSR time and an estimation of red the allowable number of single qubit operations as the function of the gate electric field at a magnetic field $B=0.1\,\text{T}$ which is parallel to the growth direction. The relaxation time calculations mainly consider the hole-phonon interactions and the details can be found in the Supplementary Information Eq.(22). Both the relaxation time and the EDSR time will depend on the spin-orbit coupling coefficients, therefore, their extrema coincide. The relaxation time increases for smaller dot sizes because that corresponds to a larger confinement energy, while both the phonon and the spin-orbit coupling terms connect the orbital ground state to higher excited states.
From Figure 4a, we can see that the Ge hole quantum dot has a long relaxation time and large Rabi ratio at dilution refrigerator temperatures. It is also useful to study the relaxation time at slightly higher temperatures, e.g. $4\,\text{K}$, at which both phonon absorption and emission must be taken into account. The phonon occupation number is given by the Bose-Einstein distribution $N=(e^{\hbar \omega/(k_\text{B} T)}-1)^{-1}$, where $N$ is the occupation number, $\omega=q v$, $q$ is the phonon wave vector and $v$ is the phonon propagation velocity, $T$ is the temperature, $k_B$ is the Boltzmann constant. More details can be found in the Supplementary Information Eq.(22), where a plot of the temperature dependence of the relaxation rate is presented as well. For $T=4\,\text{K}$, the relaxation time is $17\,\text{ms}$, suggesting that the qubit can easily be operated at this temperature.

\subsection*{Applicability and implementation}
Although we have used a simple parabolic model for the in-plane QD confinement, our conclusions are very general. Firstly, the dephasing optimal operation point will be present for potentials of arbitrary complexity (for example hut wire geometries), \cite{Katsaros.NanoLetter.Heavy-Hole.States.in.Germanium.Hut.Wires,guoguoping.APL.Measuring.hole.spin.states.of.single.quantum.dot.in.germanium.hut.wire,guoguoping.NanoLetter.Coupling.a.Germanium.Hut.Wire.Hole.Quantum.Dot.to.a.Superconducting.Microwave.Resonator}
since it is due to the fundamental interplay between the HH and LH that gives rise to the Rashba spin-orbit coupling in the HH manifold. Secondly, we have examined the possibility that the insensitivity of the $g$-factor to in-plane electric fields is an artifact of the model. We have tested three deviations from parabolicity and found that none of them exposes the qubit to dephasing by fluctuating in-plane electric fields. This implies (i) that the dot does not have to be perfectly parabolic allowing for some flexibility in the gate structure; (ii) that in-plane electric field fluctuations generally have a negligible effect on the $g$-factor, while out-of-plane electric field fluctuations cause fluctuations in the Rashba spin-orbit coupling and affect the $g$-factor, therefore it is most important to avoid the effect of the out-of-plane field; and (iii) that dephasing at the optimal operation point itself comes about primarily from higher-order terms in the electrical potential, i.e. electrical quadrupole and higher. Our results hold qualitatively in Si as well, where the spin-orbit interaction is weaker than in Ge, while $\delta$ is larger. However, the large $\delta$ and frequent failure of the Schrieffer-Wolff approximation in Si calls for a fully numerical treatment.\cite{Elizabeth.PRB.variational.analysis}

Experimentally, the configuration we describe requires a double-gated device with separate plunger gates and barrier gates allowing the number density and the gate electric-field (and spin-orbit coupling) to be controlled independently.\cite{Giordano.Advanced.Functional.Materials} The numerical estimates above suggest that, in general, a smooth and broad optimal operation point will enable the Ge hole qubit to work insensitively to the charge noise inside a large range of gate electric field accessible to experiment. Exchange-based two-qubit gates should be possible for hole QDs, and their speed depends on the values of exchange obtained, which is expected to be tunable by gates. Moreover, it is likely to simplify the coupling between the two qubits since the valley degree of freedom is absent in hole systems. However, two-qubit gate in the setup discussed here is not optimised for long distance coupling, which leads to the two-qubit gate time is of the order of microseconds for dipole-dipole interactions and hundreds of microseconds for circuit QED, limited by the Ge Luttinger parameters. They can be sped up by enhancing the spin-orbit interaction, but we defer the discussion to a future publication.

A smaller $g$-factor will lead to a smaller Rabi frequency, a smaller change in the qubit Zeeman splitting due to the spin-orbit interaction, and a shorter dephasing time but a longer relaxation time and an improved Rabi ratio. The optimal operation point will not change its location, which is determined only by the effective mass and the width of the quantum well. A larger quantum dot radius would make the confinement energy smaller, increasing the effect of the spin-orbit interaction and resulting in a faster Rabi frequency, but also shorter $T_1$ and $T_2*$. Nevertheless, the Rabi ratio decreases with increasing dot radius. Moreover, since the confinement energy decreases as the square of the radius, it is preferable to work at smaller radii to ensure the thermal broadening is overcome. Increasing $x$ increases the heavy hole - light hole splitting, leading to a reduced Rashba spin-orbit coupling and a smaller change in the qubit Zeeman splitting. The change in the Zeeman splitting will be large for smaller $x$ e.g. $0.05 - 0.10$, while at $x = 0.3$ it is essentially not noticeable. 

% \section{Summary}
We have demonstrated that electrostatically defined hole quantum dot spin qubits naturally exhibit a optimal operation point at which sensitivity to charge noise is minimized while the speed of electrical operation is maximized. The location of the optimal operation point can be determined from the width of the quantum well and the strain tensors applied.  Relaxation times are long even at $4\,\text{K}$, while dephasing is determined by higher-order terms in the expansion of the electrostatic potential due to charge defects, but are expected to allow for a large window of operation around the optimal operation point. Our results provide a theoretical guideline for achieving fast, highly coherent, low-power electrically operated spin qubits experimentally. Future studies must consider in-plane magnetic fields, which interact much more weakly with HH spins and are more complicated to treat theoretically.

\section*{Method}
\subsection*{Numerical Diagonalization}
Most of the results are obtained by theoretical analysis, we also present a numerical simulations by diagonalizing the full Hamiltonian. The full Hamiltonian is projected onto all the heavy-hole states and light-hole states and all the Fock-Darwin state.

\section*{Data Availability}
Data sharing not applicable to this article as no data sets were generated or analyzed during the current study.

\section*{Acknowledgments}
We thank Mark Friesen and Andr{\'a}s P{\'a}lyi for a multitude of engaging and educational discussions. We are also grateful to Thaddeus Ladd, Sue Coppersmith, and Andr{\text{$\grave{\text{e}}$}} Saraiva for stimulating feedback. This research is supported by the Australian Research Council Centre of Excellence in Future Low-Energy Electronics Technologies (project CE170100039) and funded by the Australian Government. 

\section*{Author Contribution}
Zhanning Wang performed all the calculations presented in this work in collaboration with Elizabeth Marcellina. James H. Cullen assisted with the theoretical calculations. Joe Salfi, Alex Hamilton and Sven Rogge contributed to the discussion of experimental implementation. Dimitrie Culcer and Joe Salfi devised the theoretical model, supervised the calculations, and wrote the manuscript.

\section*{Competing Interests}
The authors declare that there are no competing interests.

\bibliography{sample}

\section*{Figure legend and caption}

Figure 1 title: A prototype double quantum dot in a 2D hole gas. \\
Figure 1 legend:The red shaded circles represent two quantum dots confined by a set of gates. Our focus is on a single dot; two dots are shown to illustrate scaling up strategies, e.g. gates $B_2$ and $T_1$ control inter-dot tunneling.\\
Figure 2 title: The qubit Zeeman splitting.\\
Figure 2 legend: Comparison of the qubit Zeeman splitting between Schrieffer-Wolff transformation (to the third order) and exact numerical diagonalization for four different configurations. When the gate electric field is turned off, the qubit Zeeman splitting $g_0\mu_B B \approx 110\,\text{$\mu$eV}$. In all these figures, out-of-plane magnetic field is $B=0.1\,\text{T}$. We can notice that the sweet spot does not change much as a function of the quantum dot radius, but the size of the qubit Zeeman splitting will be smaller for a larger quantum dot size. In all of these plots, we have $\hbar\omega_l \gg g_0 \mu_B B$. Numerical diagonalization is the red curve, Schrieffer-Wolff method is the blue curve.(a) $\mathrm{d}=11\mathrm{nm}$, $a_0=50\mathrm{nm}$. (b) $\mathrm{d}=11\mathrm{nm}$, $a_0=60\mathrm{nm}$. (c) $\mathrm{d}=15\mathrm{nm}$, $a_0=50\mathrm{nm}$. (d) $\mathrm{d}=15\mathrm{nm}$, $a_0=60\mathrm{nm}$. \\
Figure 3 title: Dephasing time.\\
Figure 3 legend: In all plots, the quantum well width is $d=11\,\text{nm}$ and dot radius $a_0= 50\,\text{nm}$. (a) Dephasing time in the motional narrowing regime. (b) The allowable number of single qubit operations in one dephasing time in motional narrowing regime. (c) Dephasing time in the quasi-static limit.  (d) The allowable number of single qubit operations in one dephasing time in the quasi-static limit. \\
Figure 4 title: The relaxation time and the EDSR Rabi time.\\
Figure 4 legend:In all plots $d=11\,\text{nm}$, $a_0=50\,\text{nm}$, the external magnetic field is $B=0.1\,\text{T}$. The density of Ge $\rho=5.33\times10^{3}\,\text{kg}/\text{m}^3$. The phonon propagation speed along the transverse direction is $v_t=3.57\times10^{3}\,\text{m/s}$, along the longitudinal direction it is $v_l=4.85\times10^{3}\,\text{m/s}$. (a) Relaxation time and EDSR Rabi time as a function of the gate electric field. (b) The allowable number of single qubit operations in one relaxation time. 

\section{Plots}
\begin{figure}[tbp!]
\includegraphics[width=\textwidth]{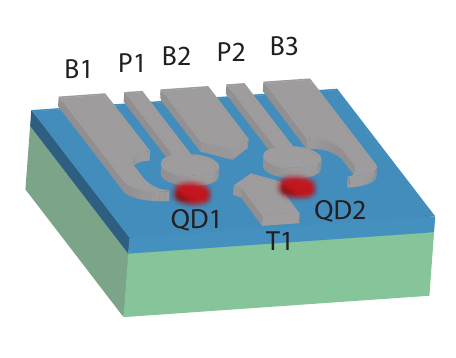}
\caption{A prototype double quantum dot in a 2D hole gas. The red shaded circles represent two quantum dots confined by a set of gates. Our focus is on a single dot; two dots are shown to illustrate scaling up strategies, e.g. gates $B_2$ and $T_1$ control inter-dot tunneling.}
\label{Figure: quantum dot in a 2D hole gas}
\end{figure}

\begin{figure}[tbp!]
\includegraphics[width=\textwidth]{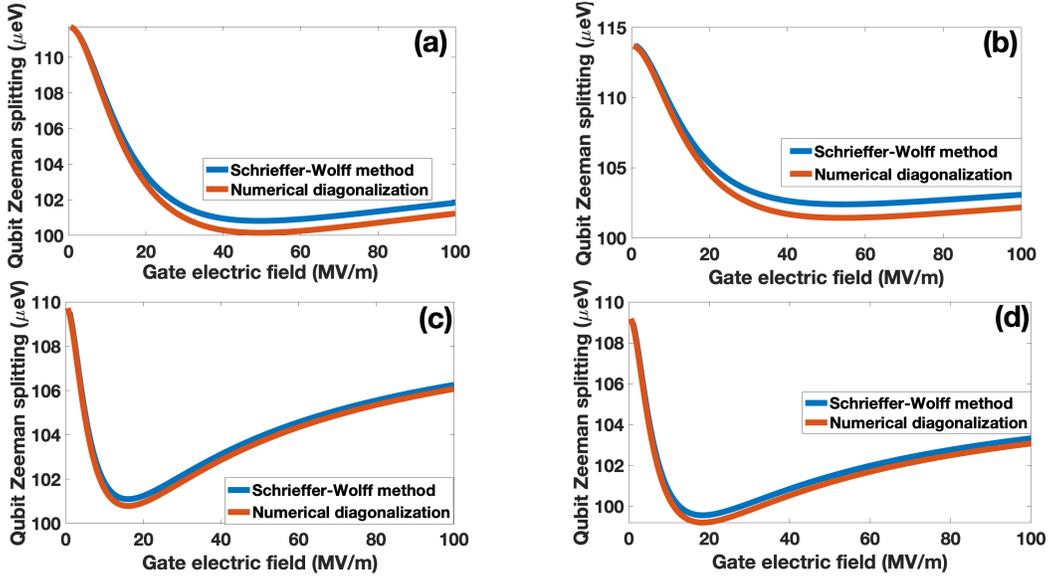}
\caption{The qubit Zeeman splitting. Comparison of the qubit Zeeman splitting between Schrieffer-Wolff transformation (to the third order) and exact numerical diagonalization for four different configurations. When the gate electric field is turned off, the qubit Zeeman splitting $g_0\mu_B B \approx 110\,\text{$\mu$eV}$. In all these figures, out-of-plane magnetic field is $B=0.1\,\text{T}$. We can notice that the sweet spot does not change much as a function of the quantum dot radius, but the size of the qubit Zeeman splitting will be smaller for a larger quantum dot size. In all of these plots, we have $\hbar\omega_l \gg g_0 \mu_B B$. Numerical diagonalization is the red curve, Schrieffer-Wolff method is the blue curve.(a) $\mathrm{d}=11\mathrm{nm}$, $a_0=50\mathrm{nm}$. (b) $\mathrm{d}=11\mathrm{nm}$, $a_0=60\mathrm{nm}$. (c) $\mathrm{d}=15\mathrm{nm}$, $a_0=50\mathrm{nm}$. (d) $\mathrm{d}=15\mathrm{nm}$, $a_0=60\mathrm{nm}$.}
\label{Figure:qubit Zeeman splitting}
\end{figure}

\begin{figure}[tbp!]
\includegraphics[width=\textwidth]{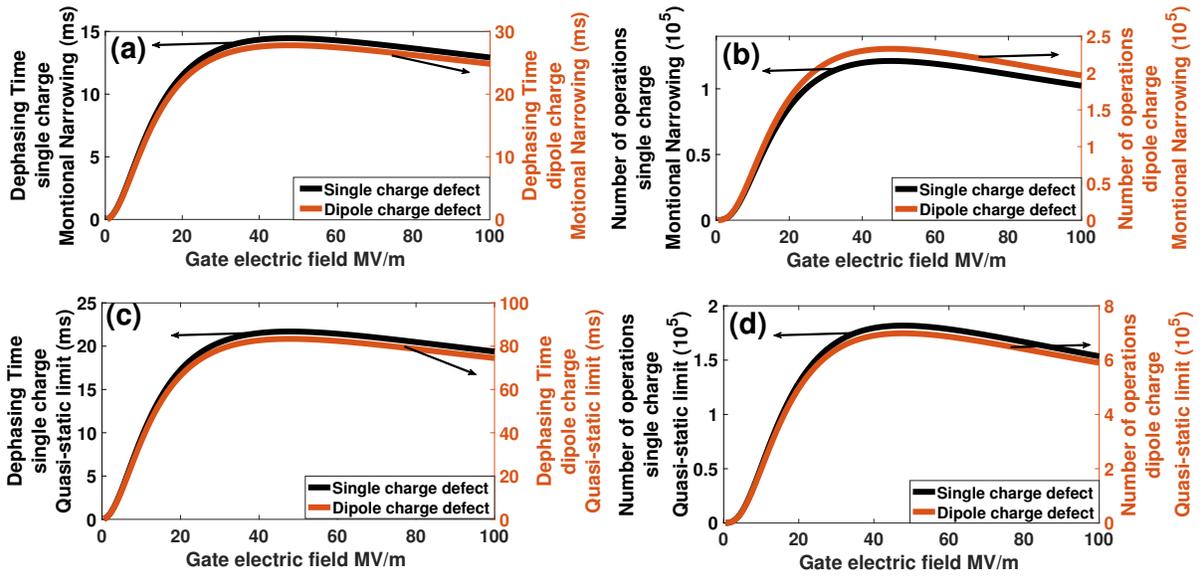}
\caption{Dephasing time. In all plots, the quantum well width is $d=11\,\text{nm}$ and dot radius $a_0= 50\,\text{nm}$. (a) Dephasing time in the motional narrowing regime. (b) The allowable number of single qubit operations in one dephasing time in motional narrowing regime. (c) Dephasing time in the quasi-static limit.  (d) The allowable number of single qubit operations in one dephasing time in the quasi-static limit.}
\label{Figure:dephasing time}
\end{figure}

\begin{figure}[tbp!]
\includegraphics[width=\textwidth]{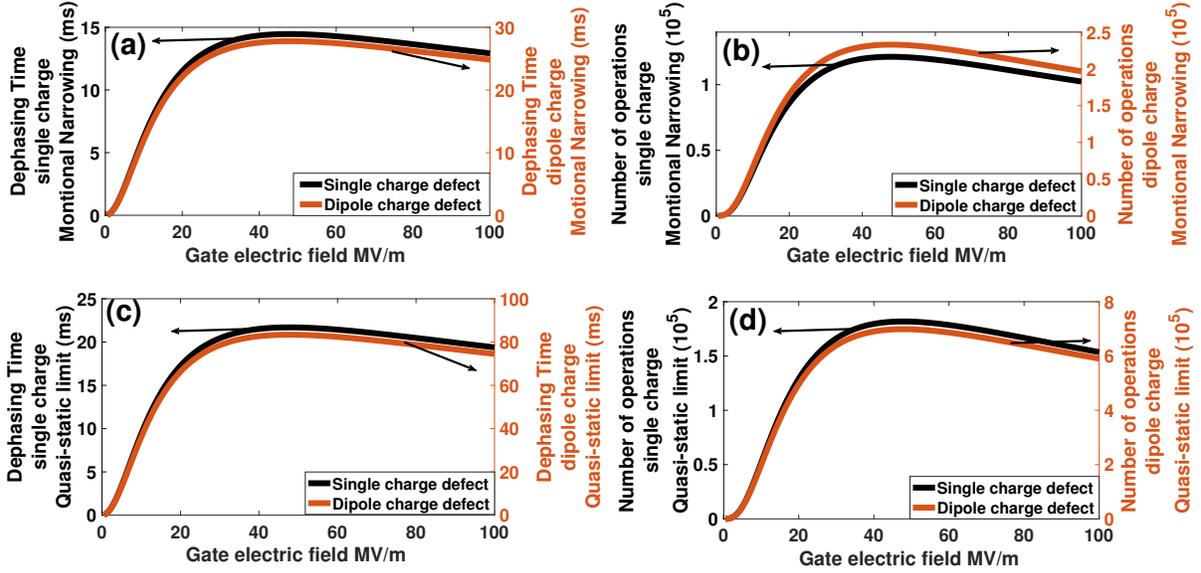}
\caption{The relaxation time and the EDSR Rabi time. In all plots $d=11\,\text{nm}$, $a_0=50\,\text{nm}$, the external magnetic field is $B=0.1\,\text{T}$. The density of Ge $\rho=5.33\times10^{3}\,\text{kg}/\text{m}^3$. The phonon propagation speed along the transverse direction is $v_t=3.57\times10^{3}\,\text{m/s}$, along the longitudinal direction it is $v_l=4.85\times10^{3}\,\text{m/s}$. (a) Relaxation time and EDSR Rabi time as a function of the gate electric field. (b) The allowable number of single qubit operations in one relaxation time.}
\label{Figure:dephasing time}
\end{figure}

\end{document}